\documentclass[prc,preprintnumbers, twocolumn, superscriptaddress, amsmath, l66amssymb]{revtex4-1}

\usepackage{graphicx}
\usepackage{dcolumn}
\usepackage{bm}
\usepackage{verbatim}
\usepackage{CJK}
\usepackage{amsbsy} 
\usepackage{microtype}
\usepackage{ulem}
\usepackage{tikz}
\usepackage{braket}
\usepackage{amsmath}
\def\BUAAISOE{School of Instrumentation Science and Opto-electronics Engineering, Beihang University, Beijing, 100191, China}
\def\BUAAHZ{Hangzhou Innovation Institute, Beihang University, Hangzhou, 310051, China}
\def\BUAARIFS{Research Institute of Frontier Science, Beihang University, Beijing, 100191, China}
\def\FUDAN{Key Lab of Nucl. Phys. \& Ion-beam Appl. (MoE), Inst. of Modern Phys., Fudan Univ., Shanghai 200433,  China}
\def\HIM{Helmholtz-Institut, GSI Helmholtzzentrum fur Schwerionenforschung, Mainz 55128, Germany}
\def\JGU{Johannes Gutenberg University, Mainz 55128, Germany}
\def\Berkeley{Department of Physics, University of California, Berkeley, CA 94720-7300, USA}
\def\NSW{School of Physics, University of New South Wales, Sydney, New South Wales 2052, Australia}

\usepackage{makeidx}
\begin{document}
\title{
New Constraints on Exotic Spin-Velocity-Dependent Interactions
}
\author{Kai Wei}
\affiliation{\BUAAISOE}\affiliation{\BUAAHZ}
\author{Wei Ji}
\email[Corresponding author: ]{wei.ji@gmail.com}
\affiliation{\HIM}\affiliation{\JGU}
\author{Changbo Fu}
\email[Corresponding author: ]{cbfu@fudan.edu.cn}
\affiliation{\FUDAN}
\author{Arne Wickenbrock}
\affiliation{\HIM}\affiliation{\JGU}
\author{Jiancheng Fang}
\affiliation{\BUAAHZ}\affiliation{\BUAARIFS}
\author{Victor V. Flambaum}
\affiliation{\JGU}\affiliation{\NSW}
\author{Dmitry Budker}
\affiliation{\HIM}\affiliation{\JGU}\affiliation{\Berkeley}
\date{\today}
\begin{abstract}

Experimental searches for new, ``fifth" forces are attracting a lot of attention because they allow to test theoretical extensions to the standard model. Here, we report a new experimental search for possible fifth forces, specifically spin-and-velocity dependent forces,  by using a K-Rb-$^{21}$Ne co-magnetometer and a tungsten ring featuring a high nucleon density. 
Taking advantage of the high sensitivity of the co-magnetometer, the pseudomagnetic field from the fifth force is measured to be $<7$\,aT.
This sets new limits on coupling constants for the neutron-nucleon and proton-nucleon interactions in the range of $\ge 0.1$\,m. The coupling constant limits are established to be $|g_V^n|<6.6\times 10^{-11}$ and $|g_V^p|<3.0\times 10^{-10}$, which are more than one order of magnitude tighter than astronomical and cosmological limits on the coupling between the new gauge boson such as Z$'$ and standard model particles.
\end{abstract}
\maketitle

\section{Introduction}

Precision measurements are powerful tools to find new physics beyond the Standard Model. For example, the discrepancies revealed by the precision measurements of the muon anomalous magnetic moment \cite{abi2021measurement,bennett2006final} and the proton radius \cite{karr2019progress,xiong2019small} have been analyzed as possible indications of ``new'' physics. Light particles, including the spin-1 boson Z$'$ and spin-0 Axion Like Particles (ALPs) were proposed to resolve these discrepancies \cite{heeck2016lepton,yan2019constraining}, 
and are also promising candidates for dark matter \cite{okada2020dark,athron2021global,co2021predictions}. 
If these particles exist, they mediate new long-range ``fifth'' forces \cite{DOB06,Fadeev2019}, and could be discovered in precision measurements.

Many experiments are conducted to search for long-range forces. 
Typical approaches include 
torsion pendulums \cite{terrano2015short}, torsional oscillator \cite{aldaihan2017calculations},
atomic magnetometers \cite{Ji2018,almasi2020new,kim2018experimental,hunter2014using}, 
 nuclear magnetic resonance \cite{tullney2013constraints,su2021search,arvanitaki2014resonantly,ledbetter2013constraints},
nitrogen-vacancy (NV) centers in diamond \cite{jiao2021experimental}, 
magnetic microscopes \cite{ren2021search}, polarized neutron experiments \cite{haddock2018search,piegsa2012limits,yan2013new}, and measurements of atomic and molecular electric dipole moments \cite{stadnik2018improved}.

Mathematically, forces between two particles, which depend on velocity, spin, and distance, can be broken down into 16 terms \cite{DOB06,Fadeev2019,moody1984new}, which provides a guide on how to search them experimentally.
 Among the 16 terms, the spin-and-velocity-dependent (SVD) terms have received extensive attention in recent years \cite{kim2018experimental,su2021search,wu2021new,jiao2021experimental,ren2021search}. 
Considering that these forces are mediated by Z$'$, the corresponding Lagrangian can be expressed as \cite{Fadeev2019}:
\begin{equation} 
\mathcal{L}_{Z'} = Z'_{\mu} \sum_\psi \bar{\psi}\gamma^{\mu}\left(g_V + \gamma^{5}  g_A  \right) \psi,
\end{equation}
where $\psi$ is the fermion field, $\gamma^5$ and $\gamma^{\mu}$ are Dirac matrices,
and $g_A$ and $g_V$ are axial and vector coupling constants.  
One of the SVD potentials, $V_{4+5}$, as being noted in Ref.\,\cite{DOB06}, can be derived from this Lagrangian \cite{DOB06,su2021search},
\begin{equation}
\begin{aligned}
V_{\rm 4+5}=&\frac{f_{4+5} \hbar^2}{8\pi m c}\left[(\hat{\boldsymbol\sigma}\cdot(\boldsymbol{v}\times\hat{\mathbf{r}})\right]\left( \frac{1}{\lambda r} + \frac{1}{r ^2} \right) e^{-r/\lambda},
\label{eq.1}
\end{aligned}
\end{equation}
where 
$ f_{4+5}=-(g_Ag_A+3g_Vg_V)/2$ is a dimensionless coupling factor,
$\hbar$ is the reduced Planck constant, 
$c$ is the speed of light, 
$\lambda$ is the force range, $\mathbf{r}$ and $\mathbf{v}$ are the relative distance and velocity between the two particles, $\hat{\sigma}$ is the spin of one fermion and $m$ is its mass.

One can write $V_{4+5}$ as
$V_{4+5}=- \boldsymbol{\mu} \cdot \boldsymbol{B}_{p}$, 
where 
$\mathbf{B}_{p}$ is the pseudomagnetic field, 
and $\mu$ is the magnetic moment of the particle.
Using bulk test material, 
the total pseudomagnetic field can be computed by integration over the volume of the material, 
\begin{align}
\begin{split}
\mathbf{B}_{p}\equiv &\frac{ f_{\perp}\hbar^2}{8\pi  m c\mu}
\int \rho_{N}(\mathbf{r})
(\boldsymbol{v}\times \hat{\mathbf{r}}) 
\left( \frac{1}{\lambda r} + \frac{1}{r^2} \right)e^{-r/\lambda}
\rm{d}\mathbf{r},
\end{split}
\label{eq.Beff.sum}
\end{align}
where $\rho_N(\mathbf{r})$ is the mass-source nucleon density at location $\mathbf{r}$ with the sensor chosen as the origin. The interaction with the neon spins are effectively averaged over the nucleons of the mass source (for example, 74 protons and 110 neutrons for a tungsten atom). 
Accordingly,
the exotic force decays exponentially with the relative distance and it is beneficial to use high-density test materials.
In fact, non-magnetic materials, such as silica (nucleon density $1.33 \times 10^{24}\,$cm$^{-3}$) and bismuth germanium oxide (BGO; nucleon density $4.3\times 10^{24}\,$cm$^{-3}$) are used as the test material in recent works \cite{jiao2021experimental,su2021search}. 
Because the possible signals are expected to be weak, a high-sensitivity magnetometer like a Spin-Exchange Relaxation Free (SERF) device \cite{allred2002high, kominis2003subfemtotesla,  kornack2005nuclear, vasilakis2009limits} is desired.

In this paper, we report a new measurement of $V_{4+5}$ by using a K-Rb-$^{21}$Ne comagnetometer \cite{quan2019synchronous},
and a tungsten test ring.  
Tungsten has a nucleon density of $1.15\times10^{25}$\,cm$^{-3}$, which is among the highest-density practically-available nonmagnetic materials. 
The sensitivity of SERF comagnetometers to pseudomagnetic fields acting on the sensor nuclei has been demonstrated to be better than 1\,fT/Hz$^{1/2} $\cite{vasilakis2009limits, chen2016spin}, which is one order of magnitude more sensitive than the ``spin-based amplifier''  demonstrated recently \cite{su2021search}. 
With these advantages, new limits on the $V_{4+5}$ neutron-nucleon and proton-nucleon interactions have been achieved. 

\section{Results}

\subsection{Experimental Setup}

\begin{figure}
\begin{center}
\includegraphics[width=9cm]{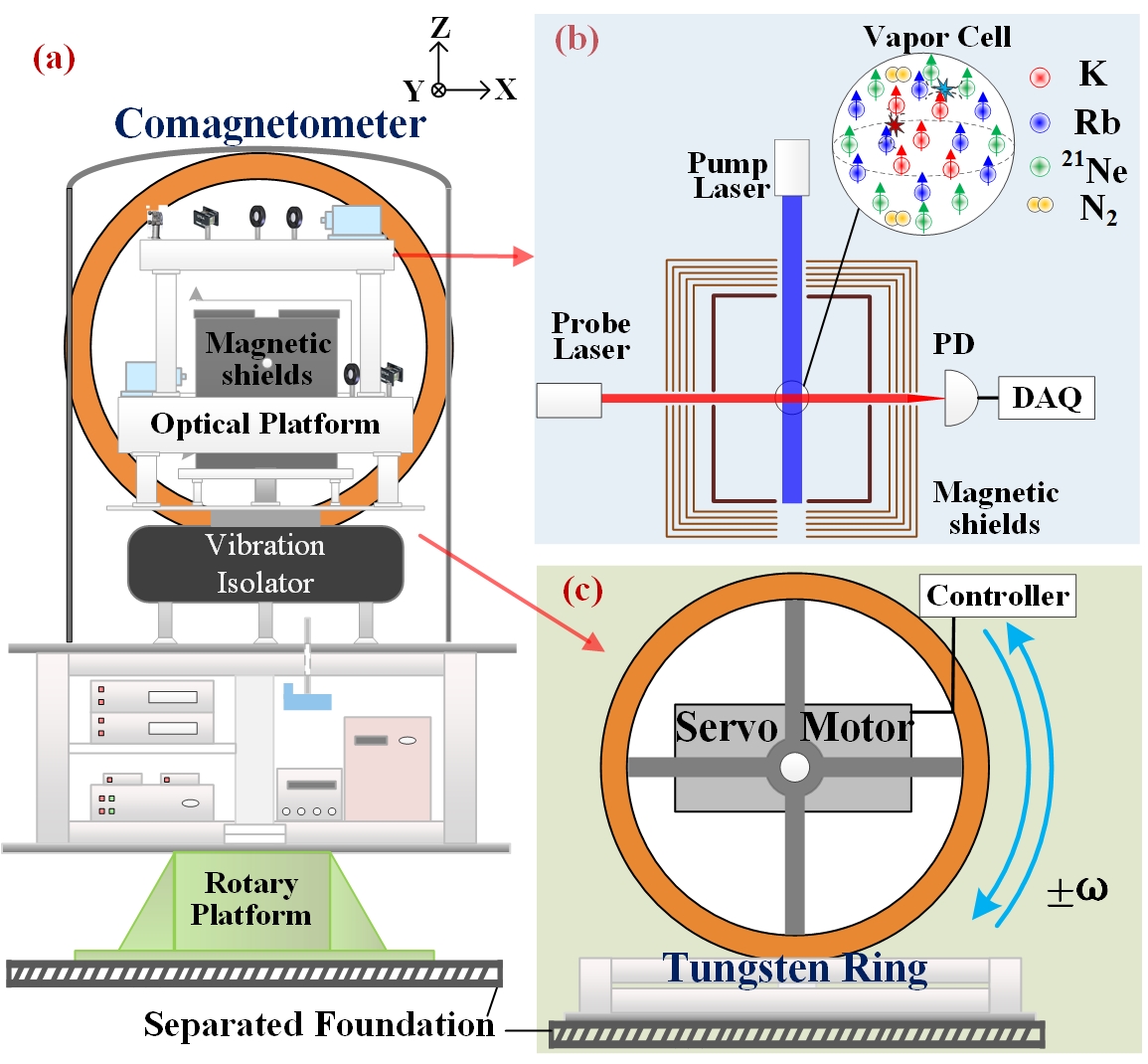}
\caption{
The experimental setup.  
(a) The structure of the whole equipment.
The tungsten ring (nucleon source) is placed at the back of the comagnetometer along $\hat{y}$ axis.
The center of the ring and the center of the comagnetometer cell are located at the same height. 
The comagnetometer is placed on a rotary platform that is isolated from the foundation of the tungsten ring.
(b) The comagnetometer setup. 
The K, Rb and $^{21}$Ne atoms are spin-polarized along $\hat{z}$ axis (vertical) by the pump laser. 
The transverse magnetic field can be measured with the probe laser.
The fluctuations of light intensities, frequencies, cell temperature are reduced with feedback. 
The external magnetic field and magnetic noise are suppressed by shielding the vapor cell in a five-layer $\mu$-metal shield and a ferrite shield.
(c) The tungsten ring.
The ring, with its duralumin support, is driven with a servo motor.
It can rotate clockwise and counterclockwise.
}
\label{Fig.Exp.Setup} 
\end{center}
\end{figure}

The experimental setup is shown in Fig.\,\ref{Fig.Exp.Setup}. 
A tungsten ring, which serves as the nucleon source, is  located behind a K-Rb-$^{21}$Ne comagnetometer. Driven with a servo motor, the ring can rotate clockwise or counterclockwise.   
The rotation axis of the ring is coaxial with the vapour cell along the $\hat{y}$ direction,
along which the comagnetometer has the highest sensitivity.
The angular position of the ring is monitored with a photoelectric encoder. 
If it exists, the pseudomagnetic field induced by the tungsten ring can be measured with the comagnetometer.

 The K-Rb-$^{21}$Ne comagnetometer is similar to that of \cite{wei2020simultaneous,chen2016spin}. Hybrid optical pumping is utilized to improve the polarization homogeneity of alkali spins and hyperpolarization efficiency of noble-gas spins, where the optically-thin K atoms are optically polarized with a circularly polarized K D1-line laser along the $\hat{z}$ axis and are used to polarize the optically-thick Rb atoms via spin-exchange (SE) collisions between K and Rb atoms. With the help of Rb and K, the  $^{21}$Ne nucleus can be polarized through the spin-exchange-optical-pumping mechanism \cite{walker1997spin, babcock2003hybrid}.
 The precession of the Rb polarization is measured via optical rotation of a linearly polarized laser beam propagating along $\hat{x}$ axis. The frequency of the laser is detuned from the Rb D1-line towards lower frequencies by about 240 GHz. Here the optically thick Rb ensemble is used for probing instead of the optically thin K ensemble, in order to improve the signal-to-noise ratio (SNR).
The precession of the $^{21}$Ne nuclei is probed with Rb atoms, and detected via optical rotation of the probe laser beam. The K-Rb-$^{21}$Ne comagnetometer is operated in the self-compensation regime to suppress magnetic noise. It is operated in the SERF regime, which results in its high sensitivity to pseudomagnetic signals \cite{kornack2005nuclear}.


 \subsection{Principle}

After zeroing the normal magnetic field, the leading terms in the comagnetometer signal in the self-compensation regime is given by \cite{almasi2020new}
\begin{equation}
\begin{aligned}
 S= K\frac{{\mathop \gamma \nolimits_e \mathop P\nolimits_z^e }}{{\mathop R\nolimits_{tot}^e }}\left( { b_y^{Ne}  - \mathop b \nolimits_y^e  + \frac{{\mathop \Omega \nolimits_y }}{{\mathop \gamma \nolimits_{Ne} }}} \right),
\label{Eq.B2V}
\end{aligned}
\end{equation}
where $\gamma_e$ and $\gamma_{Ne}$ are the gyromagnetic ratios of electrons and $^{21}$Ne nuclei; 
$P^e_z$ and $R_{tot}^e$ are the equilibrium spin polarization and transverse spin relaxation rate of alkali atoms, respectively;
$K$ is a factor to transform the $P^e_x$ to the output-voltage signal $S$; $b_y^{Ne}$ and $b_y^e$ are the exotic fields along the $\hat{y}$ axis that couple only to  $^{21}$Ne nuclei and alkali electrons, 
$\Omega_y$ is the inertial rotation rate in $\hat{y}$ axis. Deploying in Eq.(\ref{Eq.B2V}), we utilize the inertial rotations to calibrate the comagnetometer response to exotic fields, which is summarized by a scale factor $\kappa_n\equiv K \gamma_e  P_z^e / R_{tot}^e$. 

If we consider the specific coupling to fermions, 
the pseudomagnetic field on the $^{21}$Ne nuclei can be written as 
\begin{equation} 
 b^{Ne}_y=B^n_{p} \zeta_n^{Ne} +B^p_{p}\zeta_p^{Ne},
 \label{eq.B2b}
\end{equation}
where $\zeta_n^{Ne}=0.58$ and $\zeta_p^{Ne}=0.04$ are the fraction factors for neutron and proton polarization in the $^{21}$Ne nucleus, respectively \cite{brown2017nuclear,almasi2020new}, and $B_{p}^p$ and $B_{p}^n$ are the exotic fields acting on the proton and neutron, respectively. To detect $b_y^{Ne}$, a quantum nondemolition approach is used. First, the precession of $^{21}$Ne nuclear spin under the $b_y^{Ne}$ is transferred to the  Rb atoms through the Fermi-contact interaction. By measuring the precession of Rb atoms based on optical rotation, one can measure the $b_y^{Ne}$.

\subsection{Data Taking and Simulation}
The signal from the comagnetometer, as well as the angular position of the ring, are recorded with a data-acquisition (DAQ) device. Data for an individual set are taken continuously for four hours, and 24 data sets are collected in total. Between the data sets the co-magnetometer performance is optimized. The parameters of the comagnetometer, including the cell temperature, laser power, laser frequency etc., are monitored and feedback controlled throughout the experiments to ensure its stability.

\begin{figure}
\begin{center}
\includegraphics[width=9cm]{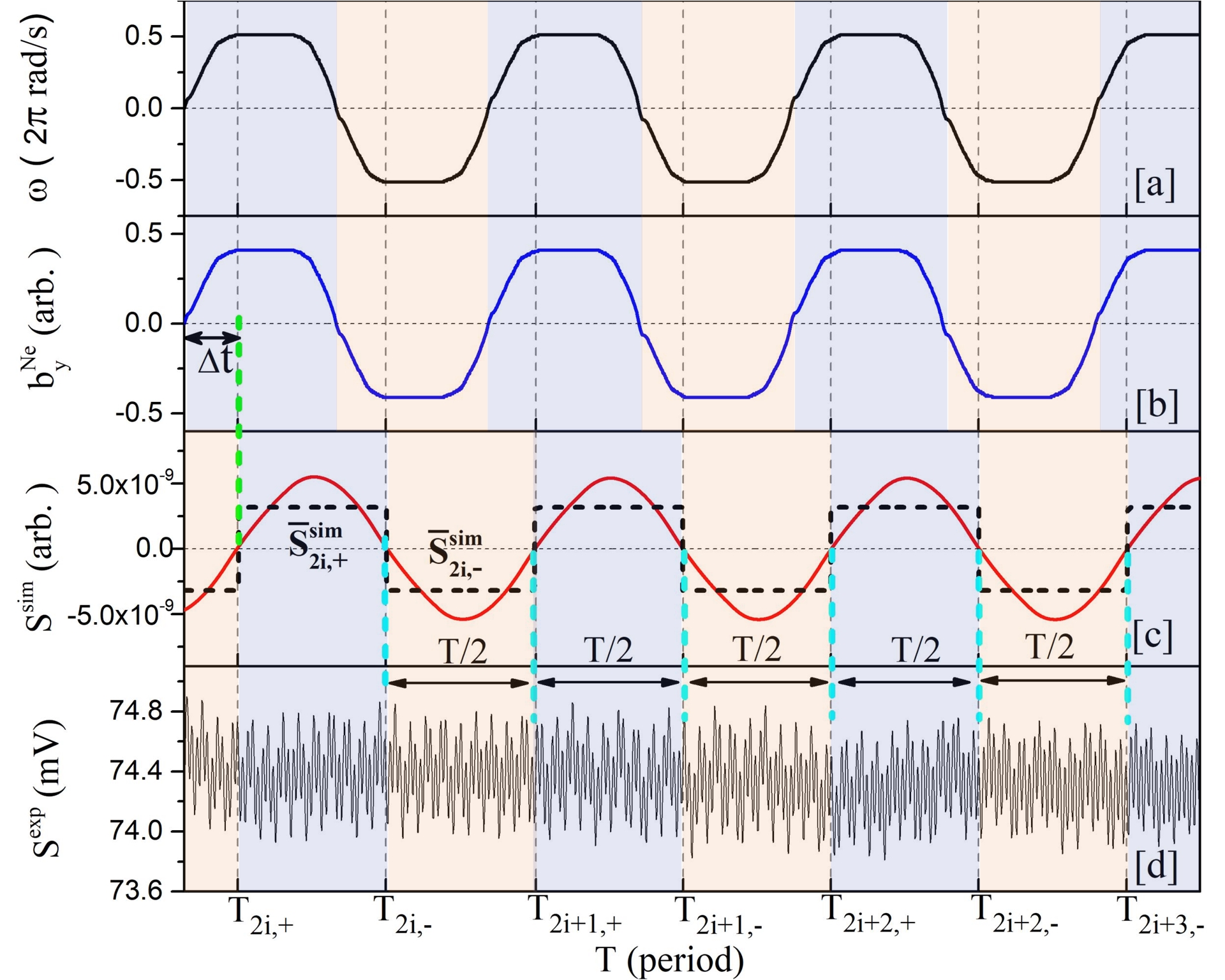}
\caption{
(a) A representative measured angular velocity of the ring, $\omega(t)$.
(b) The simulated pseudomagnetic field $b^{Ne}_{y} $. 
(c) The simulated response of the comagnetometer $S^{sim}$, which is delayed by $\Delta t$ with respect to the $b^{Ne}_{y}$. The $\bar{S}^{sim}_{2i,+}$ and $\bar{S}^{sim}_{2i,-}$ are the average of each positive and negative half cycles, respectively.
(d) The measured signal  of the SERF comagnetometer. 
}
\label{Fig.Time.Seq} 
\end{center}
\end{figure} 

 To optimize the sensitivity to the hypothetical force, it is modulated by modulating the rotation frequency of the tungsten ring. This way, the exotic signal is shifted into a frequency region with optimum noise performance of the co-magnetometer. The angular velocity of the ring $\omega^{exp}$ is measured with a high-precision photoelectric encoder and shown in Fig.\,\ref{Fig.Time.Seq}a.  The measured angular velocity is analysed using a Fast Fourier Transform (FFT) to obtain the modulation frequency and amplitude. For the displayed data the modulation frequency was measured to be 0.84\,Hz. Although $\omega^{exp}$ features a trapezoidal shape, the co-magnetometer acts like a low pass filter such that the main component of the $\omega^{exp}$-amplitude spectrum is the first harmonic of the modulation frequency, the amplitude of the third and fifth harmonic are suppressed by factor 70 and 129 respectively. 
  The maximum angular velocity is $\omega_{max}= 3.27 \pm 0.04$\,rad/s.  

The $b_y^{Ne}$ induced by the test material can be simulated using Eq.\,\eqref{eq.Beff.sum} and Eq.\,\eqref{eq.B2b}. The major input parameters for the simulation are listed in Table\,\ref{tab.inputs}, and the simulated $b_y^{Ne}$ is shown in Fig.\,\ref{Fig.Time.Seq}b. When calculating $b_y^{Ne}$, a coupling constant $f^{(0)}_{4+5}$ is assumed. In Fig.\,\ref{Fig.Time.Seq}b, it is set to be $f^{(0)}_{4+5}=1$.  


In Fig.\,\ref{Fig.Time.Seq}c, the response of the comagnetometer $S^{sim}$ is presented. This response signal is simulated based on Eq.\,\eqref{eq.Bloch} with parameters measured in the experiment. As mentioned above, because the main component of the  $b_y^{Ne}$ is at the first harmonic and the  bandwidth of the comagnetometer is narrow, the comagnetometer is mainly sensitive to the first harmonic component of $b_y^{Ne}$, which results in the approximately sinusoidal shape of  $S^{sim}$. Compared with $b_y^{Ne}$, the $S^{sim}$ has a phase shift $\Delta \phi =  
-67^{\circ}\pm 2^{\circ}$ due to the phase response of the comagnetometer (corresponding to a time delay of  $\Delta t=0.222\pm0.007\,$s).
In Fig.\,\ref{Fig.Time.Seq}d the corresponding experimental data $S^{exp}$ are shown.



\begin{table}[!h]
\begin{ruledtabular}
\caption {The experimental parameters and their error budget.}
\label{tab.inputs} 
\begin{tabular}{c c c c}  
Parameter & Value & $\Delta f^{exp}_{4+5} (\times 10^{-21})$
\footnote{The contribution  to the error budget of $f^{n}_{4+5}$ 
at $\lambda=10$ m} \\
W-Al ring R (m) &$0.475\pm 0.001 $ &$<0.001$\\
 W-Al ring M (kg)  &$15.38\pm 0.05$ & 0.001\\
Ring to cell center X (m) & $0.478\pm 0.002$& 0.002 \\
Ring to cell center Y (m) & $0.002\pm 0.002$& $<0.001$\\
Ring to cell center Z (m) & $0.002\pm 0.002$&$<0.001$\\
Modulation freq. (Hz)& $0.83681 \pm 0.00001$& $<0.001$\\ 
$\omega_{max}$ (rad/s) & $3.27\pm 0.04$\ & 0.004\\
Phase shift ($\deg$) & $-67^{\circ}\pm 2^{\circ}$ & $<0.001$\\
Calibrated $ \kappa_n$ ($\mu$V/fT) & $ 1.67 \pm 0.05$ &0.011\\ 

\hline
Final $f^{exp}_{4+5}$ & $0.36$
 & $\pm$ 0.01 (syst.)\\
$(\lambda=10\,$ m) &  & 
$\pm 1.07$ (stat.)
\footnote{
Error contribution from the uncertainties of the parameters listed above.
} 
\end{tabular} 
\end{ruledtabular}
\end{table}

\subsection{Limits on exotic forces}

\begin{figure}  
    \centering
    \includegraphics[width=9cm]{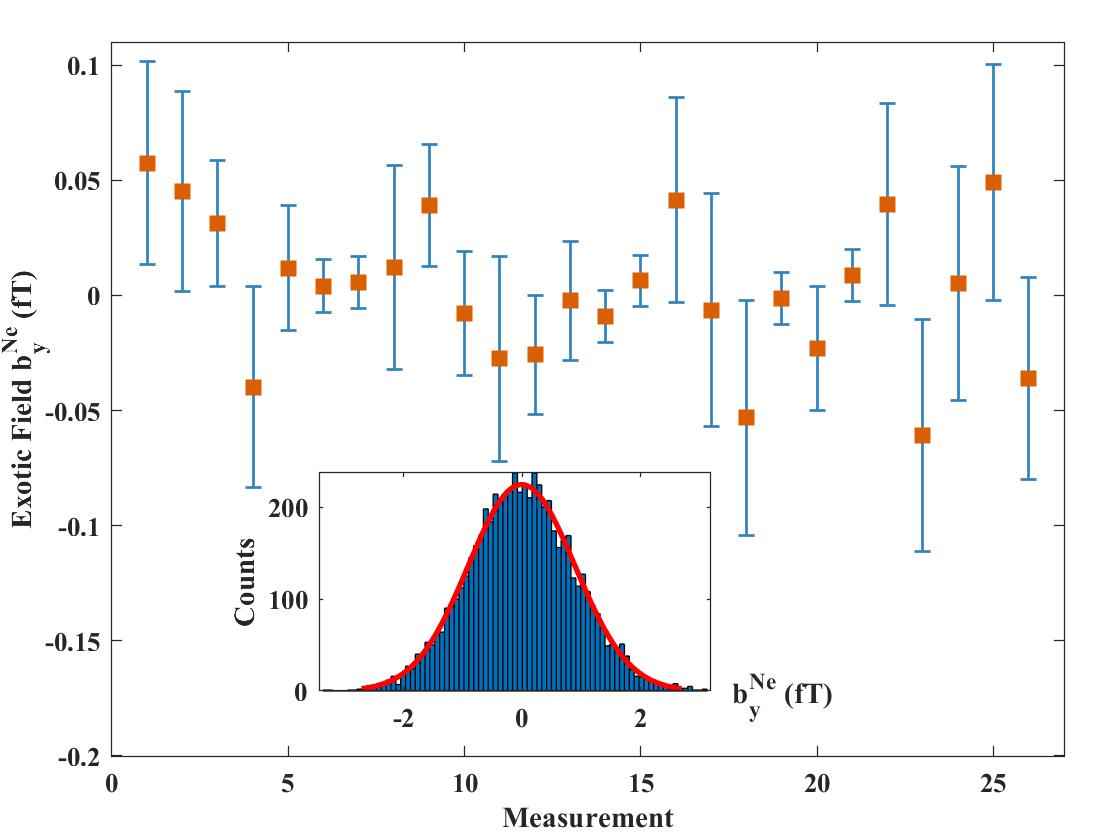}
\caption{
Experimental results of the $b^{Ne}_y$.
Each point represents an average of a 4-hours data set, $\bar b$$^{Ne}_y$. 
Histogram of $\bar b$$^{Ne}_y$ is shown on the inset, with the red curve being a Gaussian fit.
The exotic field $b^{Ne}_y$ is measured to be $(2.4\pm7.1)$\,aT.
}
\label{fig:total_data}
\end{figure}



\begin{figure}
\begin{center}
\includegraphics[width=9.cm]{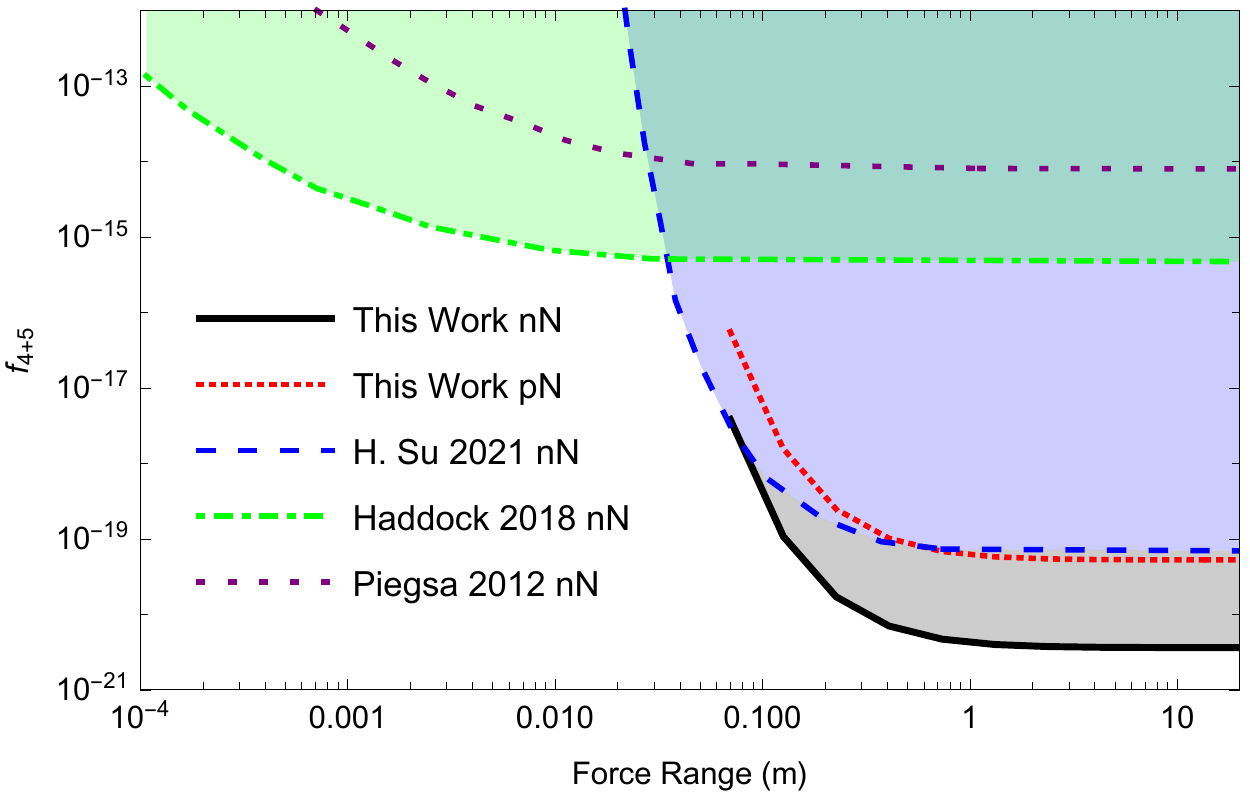}
\caption{
The experimental limits on $f_{4+5}$. 
The ``nN"  denotes the coupling between neutron and nucleon, and ``pN" for proton and nucleon.
The blue dashed line, ``H.Su 2021", is from Ref.\,\cite{su2021search}, 
the green dash-dot line, ``Haddock 2018", is from \cite{haddock2018search}, 
the yellow doted line, ``Piegsa 2012",  is from \cite{piegsa2012limits}. 
The black solid line 
and red doted line represents our new results for ``nN" and ``pN" respectively.
}
\label{Fig.Limits45.Setup} 
\end{center}
\end{figure} 

The coupling constant can be found by
\begin{equation}
  f_{4+5}=f_{4+5}^{(0)}\frac{\bar{S}^{exp}}{\bar{S}^{sim}},\end{equation} 
where the definitions of $\bar{S}^{exp}$ and $\bar{S}^{sim}$ can be found in Materials and Methods, Eq.\,\eqref{eq.Quad_BG}.

  A typical $\bar{S}^{exp}$ distribution, deduced from  4-hour data,  is shown in the inset of Fig.\,\ref{fig:total_data}.  Note that the  proton fraction of polarization in the Rb atom is $\zeta_p^{Rb}=0.31$ \cite{kimball2015nuclear}, but the Rb atoms' energy sensitivity is three orders of magnitude smaller than that of $^{21}$Ne \cite{almasi2020new,padniuk2022response}. Thus we only consider the proton spin in $^{21}$Ne. Using all the 104-hour data, the pseudomagnetic field $b_y^{Ne}$ is measured to be $(2.4 \pm 7.1) $\,aT. This results in new limits on the dimensionless coupling factor of the new force $|f_{4+5}|\le 2.21\times 10^{-21}$ with 95$\%$ confidence level at the force range of 10 m. The limits on the coupling between the source nucleon with the $^{21}$Ne neutron and neutron are set to be $|f^n_{4+5}|=|f_{4+5}|/\zeta^{Ne}_n\le 3.8\times 10^{-21}$ and $|f^p_{4+5}|=|f_{4+5}|/\zeta^{Ne}_p\le 5.5\times 10^{-20}$ respectively. Our experiment results and a comparison with the results in the literature are shown in Fig.\,\ref{Fig.Limits45.Setup}. Our result on the coupling with neutron is more than an order of magnitude better than that of previous works. The limits on exotic coupling with proton is set by this work for the first time to the best of our knowledge. 

It may be useful to consider specific cases for coupling, using the fact that $f_{4+5}= \frac{1}{2} g_A g_A -\frac{3}{2} g_V g_V$. 
Assuming $g_Ag_A=0$, we have
$|(\zeta^{Ne}_p g^p_V+\zeta_n^{Ne} g^n_V)(\zeta_p^{W} g^p_V+\zeta_n^{W} g^n_V)|\le 1.47\times 10^{-21}$, where $\zeta_n^W=0.588$ and $\zeta_p^W=0.412$ are the neutron and proton mass contribution in the W-Al mass source. If we assume $g^p_V=0$, we get a limit on $|g^n_V|\le6.6\times 10^{-11}$. Conversely, if we assume $g^n_V=0$, we get $|g^p_V|\le3.0\times 10^{-10}$. The coupling limit $|g^p_A|\le 9.0\times 10^{-10}$ and $|g^n_A|\le 2.0\times 10^{-10}$ is also set using the same method. 


\begin{figure}
\begin{center}
\includegraphics[width=9.cm]{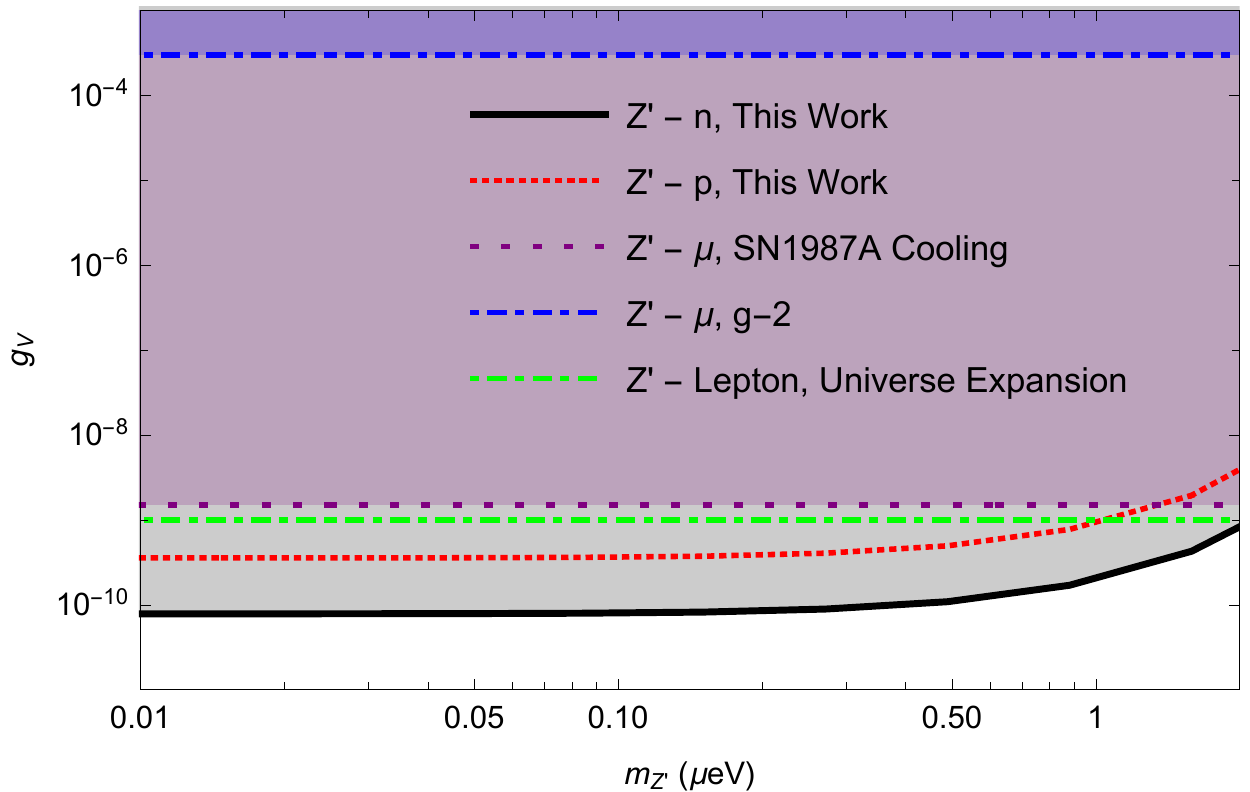}
\caption{
The experimental limits on the vector coupling constant $g_V$ of the Z$'$ particle. 
The black solid line and red dotted line are set by this work by searching for the fifth force. 
The blue dot-dashed line is set by the $g-2$ experiment\,\cite{croon2021supernova,escudero2019cosmology,tanabashi2018review}, the green dot-dashed line is set by the neutrino production in the expansion of the  early universe\,\cite{escudero2019cosmology},
the purple dashed line, is from the analysis of the supernova SN1987 cooling\,\cite{croon2021supernova}.
}
\label{Fig.Limits.gV} 
\end{center}
\end{figure} 

A comparison of the limits on $g_V$, the vector coupling constant between Z$'$ and standard model particles, between this fifth-force result and other results including the cosmology and astronomy is shown in Fig.\,\ref{Fig.Limits.gV}. The `Z$'$-Lepton Universe Expansion' line is excluded by the effective number of neutrino species $\Delta N_{nef}\approx 0.2$ in the early universe\,\cite{escudero2019cosmology}. The authors of this work assume that Z$'$ can decay to neutrinos and affect the expansion of the universe. This model can relax the 3$\sigma$ tension of Hubble constants, i.e., the discrepancy between local measurements and temperature anisotropies of the cosmic microwave background\,\cite{aghanim2021erratum,riess2018milky,escudero2019cosmology}. The `Z$'$-$\mu$, $g-2$' is excluded by the muon $g-2$ experiment\,\cite{croon2021supernova,escudero2019cosmology,tanabashi2018review}. Note that the anomalous magnetic moment of muon can also be used to search for the spin-0 boson, such as axion like particles\,\cite{yan2019constraining}. The `Z$'$-$\mu$, SN1987A Cooling' is excluded by the supernova SN1987A, assuming the new gauge boson Z$'$ decreases the cooling time\,\cite{croon2021supernova}. Our results represent more than one order of magnitude tighter constraints than previous ones.

\section{Discussion and conclusions}

The main advantage of this experiment compared to that in \cite{su2021search} are attributed to the high nucleon density of the nucleon source and the ultrahigh sensitivity of the comagnetometer. Tungsten has nucleon density approximately four times that of BGO, and our comagnetometer sensitivity is approximately one order of magnitude better than the $^{129}$Xe based magnetometer used in Ref.\,\cite{su2021search}.

The comagnetometer can also be used to search for many terms of spin-spin-velocity-dependent forces if one can use a spin-polarized source, such as a pure-iron shielded SmCo$_5$ electron-spin source \cite{Ji2018}. For four terms of the spin-spin-velocity-dependent forces, new limits on electron-neutron and electron-proton interaction can be set in the force range $\ge$1\,cm \cite{ji2017searching}, which would complement the results of \cite{hunter2014using} for interaction ranges exceeding 1\,km.

The techniques used in this work can be used to search for a broad class of beyond-standard-model particles \cite{moody1984new,DOB06,Fadeev2019}. The search for such particles is motivated, among other things, by the attempt to understand the composition of the dark sector (dark matter and dark energy). However, similar to other fifth-force searches, it does not, in any way, rely on specific local dark-sector properties, e.g., the local dark-matter density. 

If the exotic-force mediator is a Z$'$-boson, the interactions could violate parity, however, the current experiment does not exploit parity violating effects. Indeed, the interaction in Eq.\,\eqref{eq.1} is P- and T-even.

In conclusion, in this work, we have searched for exotic spin- and velocity-dependent interactions and, for the force range larger than several centimeters, improved the limits on the mass interactions with neutrons by more than an order of magnitude, while also setting a stringent limit on mass interactions with protons.
This result demonstrates that the fifth-force approach is competitive in terms of sensitivity to new bosons with the cosmological and astronomical searches. Z$'$ can also be searched through pseudovector  parity-violation couplings\,\cite{antypas2019isotopic,Fadev2022Pseudovector}.


\section{Materials and Methods}


\subsection{Modulation of the exotic force}
The angular velocity $\omega(t)$ of the tungsten ring is designed to have the following pattern: accelerating period        [$\omega(t)= \alpha t$,     $(n-\frac{2}{14})T\le t\le (n+\frac{2}{14})T$]; running positively period  [$\omega(t)= \omega_{max}$, $(n+\frac{2}{14})T\le t\le (n+\frac{5}{14})T$]; decelerating period        [$\omega(t)= \omega_{max}-\alpha t$,     $(n+\frac{5}{14})T\le t\le (n+\frac{9}{14})T$]; 
and running negatively period  [$\omega(t)= -\omega_{max}$, $(n+\frac{9}{14})T\le t\le (n+\frac{12}{14})T$],
where $\alpha$ is the  acceleration rate  and $T$ is the period for each modulation cycle. Due to the large moment of inertia of the ring, the maximum acceleration rate of the servo motor is limited to $\alpha=6 \pi$\,rad/s$^2$.  
\subsection{ Calibration}

\begin{figure}
\begin{center}
\includegraphics[width=9cm]{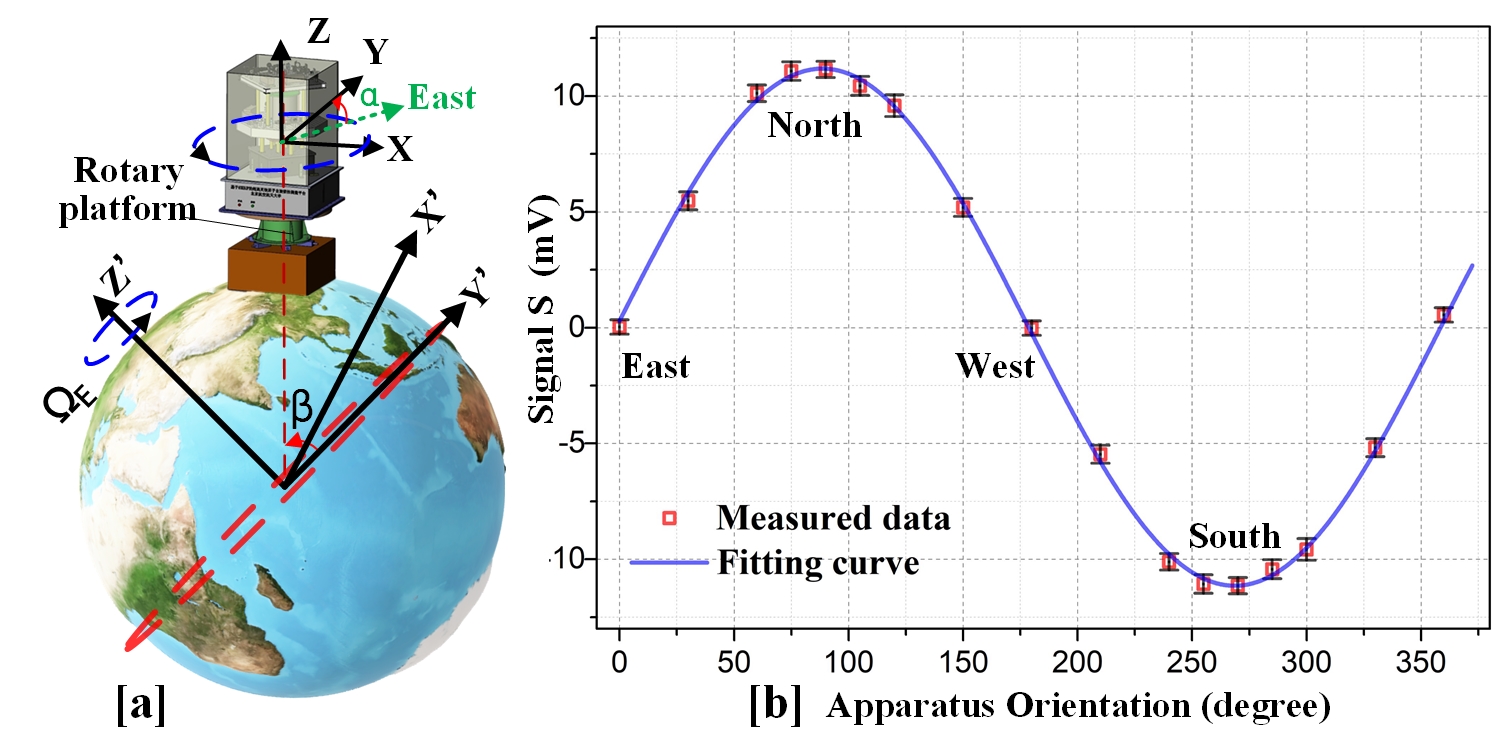}
\caption{
The calibration of the comagnetometer by the Earth rotation. $\beta$ is the latitude of the lab, $\alpha$ is the orientation angle of $\hat{y}$ with respect to the east direction in the horizontal plane. The measured dots are fitted by a sinusoidal function with R-square better than 0.99.}
\label{fig.Ear.Rota.Cal} 
\end{center}
\end{figure}

The Earth rotation speed $\Omega_E=7.292\times10^{-5}$\,rad/s is used to calibrate the system \cite{brown2010new}.
Mounted  on a  high-precision rotary platform,  the apparatus can rotate in the horizontal plane as shown in Fig.\,\ref{fig.Ear.Rota.Cal} [a].
The $\Omega_y$ in Eq.\,(\ref{Eq.B2V}) can be written as
$\Omega_y=\Omega_E^{h} sin (\alpha)$,
where $\Omega_E^{h}= \Omega_E cos(\beta)$  is the projection of $\mathbf{\Omega}_E$ in the horizontal plane, $\beta =39.983^{\circ}$ is the latitude of the laboratory, and $\alpha$ is the relative  azimuth angle of the sensitive $\hat{y}$ axis of the comagnetometer. Therefore, by fitting the measured signals with 
$S(\alpha)=\kappa_n \Omega_E^{h} sin (\alpha)/\gamma_n $, 
the scale factor $\kappa_n$ can be obtained. As shown in  Fig.\,\ref{fig.Ear.Rota.Cal} [b], the scale factor at this near-DC frequency is measured to be  $\kappa_n$(DC)=$(4.18 \pm 0.07) \times 10^{-6} $\,V/fT with a R-square factor (goodness of fit) better than 0.99. 

 In the self-compensation comagnetometer, the frequency responses to normal magnetic fields and inertial rotations ($B_{x/y}$ and $\Omega_{x/y}$) and exotic fields ($b_y^{Ne}$ and $b_y^{e}$) are all determined by four parameters, which are the Fermi-contact-interaction fields between noble-gas atoms and alkali atoms ($\lambda  M^{e}_0  {P}^e_z$ and $\lambda  M^{n}_0  {P}^n_z$)  \cite{romalis1998accurate}, and the transverse relaxation rates ($1/T_{2e}$ and $1/T_{2n}$). These four parameters are independently measured to be $110.6\pm2.2$\,nT, $579.4\pm2.1$\,nT, $3715\pm445$\,1/s, and $0.063\pm0.007$ 1/s, respectively  \cite{wei2020simultaneous}. Particularly, the noble-gas nuclear spins and alkali electron spins are strongly coupled around the self-compensation point, such that the damping rate of nuclear spins is sped up by electron spins, while electron spins are slowed down  \cite{kornack2002dynamics}. The damping rate of nuclear spins is measured to be $2\pi \times (0.907\pm 0.006)$\,1/s by fitting the damping oscillation response to step magnetic field excitation, which is significantly larger than the $1/T_{2n}$ owing to the coupling between nuclear spins and electron spins.

\subsection{Simulation}

The spin dynamics of the comagnetometer could be described by the coupled Bloch equations \cite{kornack2005nuclear, wei2020simultaneous}: 
\begin{eqnarray}\label{eq.Bloch}
   \frac{{\partial  {\bf{P}}^{e} }}{{\partial t}} &=& \frac{{ \gamma _e }}{Q}\left({\bf{B}} + \lambda  {{M}}^n_0  {\bf{P}}^n + {\bf{b}}^{e} +{\bf{\Omega}}\frac{Q }{\gamma _e } \right) \times  {\bf{P}}^e  \nonumber\\
   &&+ \frac{P^e_{z0}\hat{z} -  {\bf{P}}^e }{Q\{  T_{1e} , T_{2e} ,T_{2e}  \}},\nonumber \\
   \frac{{\partial  {\bf{P^n}} }}{\partial { t}} &=&  \gamma _{Ne} \left({\bf{B}} + \lambda  M^e_0  {\bf{P}}^e + {\bf{b}}^{Ne} + \frac{\bf{\Omega }}{\gamma _{Ne}} \right) \times {\bf{P}}^{\bf{n}}  \nonumber\\
   &&+ \frac{P^n_{z0}\hat{z} -{\bf{P}}^n }{\{  T_{1n} , T_{2n} ,T_{2n}  \}},
\end{eqnarray}
where  $Q$ is the slowing-down factor arising from spin-exchange collisions and hyperfine interaction \cite{savukov2005effects},
$\bf{P^e}$ and $\bf{P^n}$ are the spin polarizations of alkali electron  and  $^{21}$Ne nuclear, respectively. 
$\bf{B}$ and $\bf{\Omega}$  are external magnetic field and inertial rotation. The Fermi-contact interaction between alkali atoms and $^{21}$Ne atoms can be described by an effective magnetic field $\lambda  M^{e,n}_0  {\bf{P}}^{e,n}$, where $M^n_0$ ($M^e_0$) is the maximum magnetization of $^{21}$Ne nucleon (alkali electron) \cite{romalis1998accurate}. For a uniformly spin polarized spherical cell, $\lambda = 8\pi\kappa_0/3$, where $\kappa_0$ is the enhancement factor. $ T_{1e}$ and $ T_{2e}$ are the longitudinal and transverse relaxation rates for alkali electron spin, respectively, and $T_{1n}$ and $T_{2n}$ are the longitudinal and transverse relaxation times for the $^{21}$Ne nucleon spin.

\begin{figure}
\begin{center}
\includegraphics[width=8.5cm]{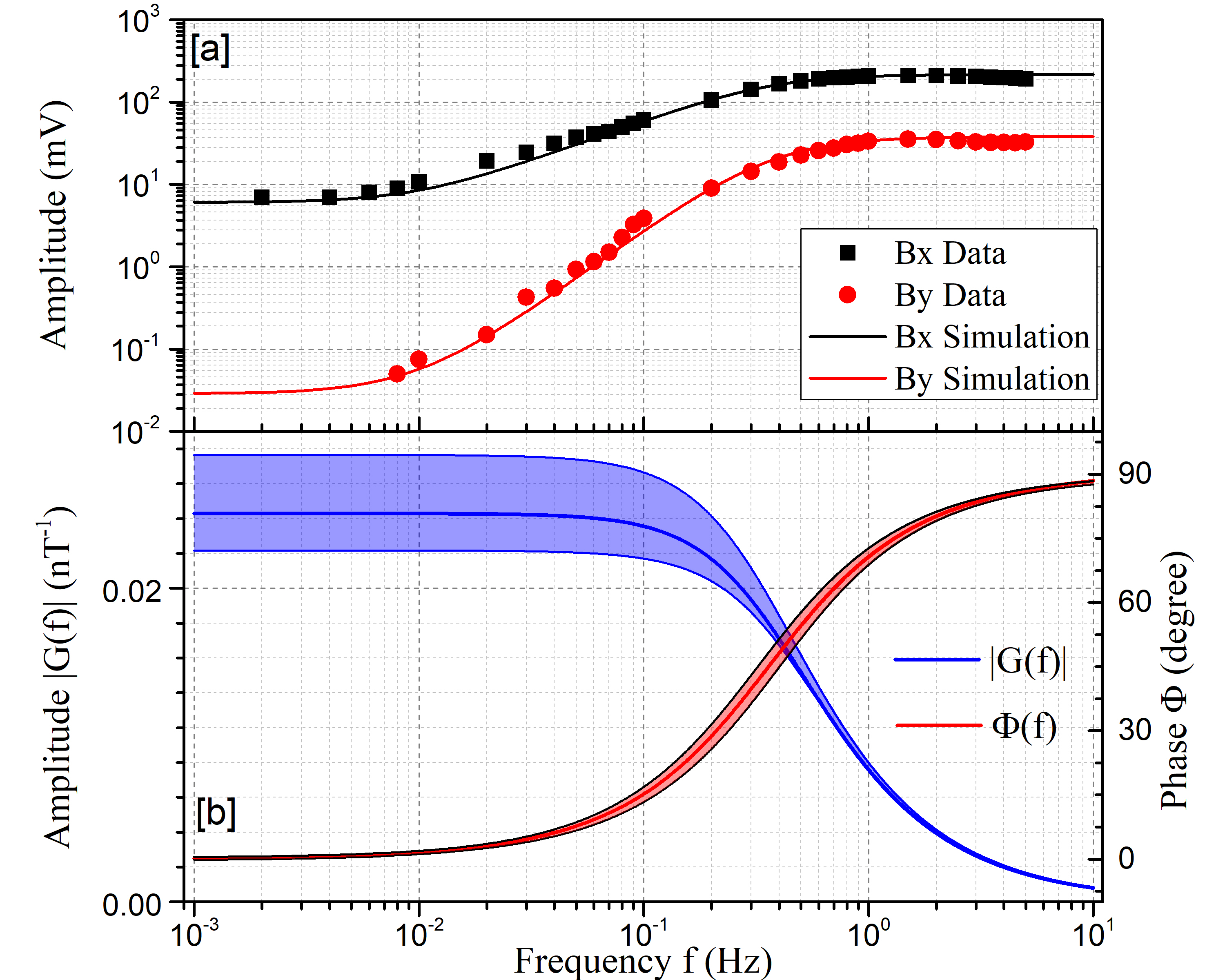}
\caption{
[a]  A comparison of the measured frequency responses to normal magnetic field and the simulated results. The simulated results are calculated with Eq.(\ref{eq.Bloch}) based on  measured parameters ($\lambda  M^{e}_0  {\bf{P}}^e_z$, $\lambda  M^{n}_0  {\bf{P}}^n_z$, $T_{2e}$ and $T_{2n}$) with only one free parameter to describe the scale.  The measured data agrees well with the simulation of the Bloch equation Eq.(\ref{eq.Bloch}), which confirms the validity of the simulation.
[b]  The simulated amplitude-frequency response (blue) and phase-frequency response (red) to exotic field b$^{Ne}_y$. The uncertainty bands of the phase and amplitude are calculated based on the uncertainties of the measured parameter.}.
\label{fig.Fre.Res} 
\end{center}
\end{figure}

As shown in Fig.\,\ref{fig.Fre.Res} [a], to further verify the validity of the parameters  $\lambda M^{e}_0 {P}^e_z$, $\lambda  M^{n}_0  {P}^n_z$, $T_{2e}$, and $T_{2n}$) and the frequency response model, the frequency responses to $B_y$ and $B_x$ are measured and compared with the simulated results with Eq.\,(\ref{eq.Bloch}) based on these four parameters with only one free parameter to describe the scale.  The measured signals are consistent with the simulated results.
Therefore, the response of the comagnetometer $S^{sim}$ to the exotic field $b_y^{Ne}$ can be simulated by solving the Bloch Eqs.\,\eqref{eq.Bloch} \cite{padniuk2022response} with these verified parameters. The simulation result of the amplitude and phase response to $b_y^{Ne}$ is shown in Fig.\,\ref{fig.Fre.Res} [b]. The results are further used to correct the scale factor $\kappa_n$ and the phase retardation $\phi$. For the modulation frequency at 0.83681(1) Hz, the amplitude response is $0.40\pm0.01$ of that in DC. Meanwhile, the phase shift $\Delta \phi$ is $-67^{\circ}\pm 2^{\circ}$. The corrected scale factor is $\kappa_n (\textrm{0.83 Hz})= (1.67 \pm 0.05) \times 10^{-6}$\,V/fT. In addition, the calculated bandwidth in Fig.\,\ref{fig.Fre.Res} [b] is consistent with the measured sped-up damping rate of noble-gas nuclear spins.

\subsection{Experimental Setup}
The test material includes the tungsten ring  and its duralumin holder, both can generate the exotic force. The tungsten has nucleon density of $1.15\times10^{25}$\,cm$^{-3}$, while the duralumin has a nucleon density of $1.67\times 10^{24}$\,cm$^{-3}$.
The inner radius of the tungsten ring is $R=(0.475 \pm 0.001)$\,m.
The ring is composed of ($11.20 \pm 0.05$)\,kg of tungsten wires wrapped on a  ring-shape duralumin support, which has a total mass of  $(4.18\pm 0.01)$\,kg. The height of the center of the ring is the same as the height of the center of the vapour cell, and  the distance between the center of the ring and the center of the comagnetometer cell is $D_c=(0.478 \pm 0.003)$\,m. 
Four duralumin supporting rods are installed to connect the aluminium-alloy spindle and the ring-shaped support. 
The total mass of these support rods is $(3.73\pm 0.01)$\,kg. Based on the simulation, the tungsten-duralumin (W-Al) ring part generate 90\% of the exotic force, while the other moving parts in the device generate the rest.

The K-Rb-$^{21}$Ne comagnetometer works as the sensor. A spherical glass cell  is filled with  2500\,torr of $^{21}$Ne ($70\%$ isotope enriched) and 54\,torr of N$_2$, as well as a small droplet of K and Rb atoms. The cell is heated to 200$^\circ$C with a proportional-integral-differential (PID) controlled AC electric heater, resulting in temperature fluctuations of less than 5\,mK. The heating current is modulated at 100\,kHz to reduce low-frequency magnetic interference and the heating wire is double-layer printed in a flexible thin film to further reduce the magnetic field from the heating wires. At 200$^\circ$C, the number densities of K and Rb atoms are around 8$\times 10^{12}$\,cm$^{-3}$ and 8$\times 10^{14}$\,cm$^{-3}$, leading to a density ratio of about 1/100. The cell and the heater are enclosed inside a water-cooled vacuum chamber to improve the temperature stability as well as reduce the air convection.   The cell is placed inside a five-layer $\mu$-metal magnetic shield to reduce the ambient magnetic field. Additionally, a Mn-Zn ferrite shield, which has high resistivity \cite{kornack2007low}, is placed in the innermost layer to minimize the magnetic noise from the shielding material. The shields provide an overall shielding factor better than 10$^8$ at low frequency. 

Linearly polarized light detuned towards lower frequencies from the D1 line of Rb is used  to probe the spin precession of Rb atoms, originating from the potential precession of  $^{21}$Ne nucleon magnetization due to the exotic field. Furthermore, to avoid low-frequency noise, the probe light is modulated with a photo-elastic modulator (PEM) and the signal is demodulated with a lock-in amplifier. The intensities of pump light and probe light are actively stabilized with liquid crystal variable retarders using PID feedback control to suppress intensity-related noise. Meanwhile, to reduce light-frequency noise, the probe laser is stabilized with a two-stage temperature controller, while the frequency of pump light is locked via saturated absorption stabilization.

Because we use the hybrid optical pumping to improve the polarization homogeneity, higher pump light intensity is required. Pump laser light (1.3\,W) from a tapered amplifier is coupled into an optical fiber to clean the spatial mode. The light is intensity stabilized with a liquid crystal modulator and is expanded to cover the $12$\,mm diameter cell. The intensity of the incident pump light is about 564\,mW/cm$^2$. The spin ensemble is typically polarized for at least five hours to reach the quasi-steady state  with a leading field along $\hat{z}$ of about 900\,nT before we start to execute the field-zeroing procedure. After degaussing with a 60\,A induction degaussing device, the residual magnetic fields and field gradient at the center of the magnetic shields are measured with a fluxgate magnetometer to be smaller than 1\,nT and 1\,nT/cm, respectively. These are further reduced using compensation coils. The self-compensation field and other magnetic fields are periodically corrected using the field-zeroing procedures. Because temperature, laser intensity, light frequency, air convection and the vibration are carefully controlled, the drift of the self-compensation field is smaller than 0.5\,nT in one day. The calibration of scale factor is checked periodically. The photon shot noise is  calculated  to be smaller than 0.1\,fT/Hz$^{1/2}$, while the  measured background noise of the probe light exceeds 0.3\,fT/Hz$^{1/2}$. The electronic noise of the photodetector and lock-in amplifier is smaller than 0.1\,fT/Hz$^{1/2}$. To reduce the background noise, the optical polarization of probe light is adjusted with a quarter-wave plate to compensate the birefringence of vacuum windows and the cell based on the response to magnetic field modulation. To operate the system in a steady state, the comagnetometer is continuously operated for several weeks.

Sensitive searches for dark matter and exotic fields based on atomic spins are often limited by the intrinsic magnetic noise generated by the innermost magnetic shield itself. We use the low-noise Mn-Zn ferrite shield for the innermost shielding layer. Different kinds of Mn-Zn ferrite with different relative permeability and geometry were tested.  After 
comparing, we chose the ferrite shield whose real and imaginary components  of the relative permeability were measured to be $\mu '/\mu_0=6308$ and $\mu ''/\mu_0=45$, respectively. The length, inner diameter and  thickness of the cylindrical ferrite shield are 22\,cm, 11.4\,cm and 1.3\,cm, respectively. Although the AC electric heater is enclosed inside a vacuum chamber whose outer surface is wrapped with water-cooled tube,  the actual temperature of the inner surface of the ferrite shield is still higher than room temperature and is measured to be 47$^\circ$C. Using the above parameters, the magnetic noise of the ferrite shield is calculated to be 2.5$f^{-1/2}$\,fT \cite{kornack2007low} ($f$ is the frequency in Hz), which is 2.8\,fT/Hz$^{1/2}$  at the modulation frequency of the nucleon source (0.84\,Hz). Since the comagnetometer is operated close the self-compensation regime, this further suppresses the magnetic noise (see the response curve in Fig.\,\ref{fig.Fre.Res}. The overall noise of the comagnetometer is translated into the effective pseudomagnetic field noise level, which is measured to be 1.5\,fT/Hz$^{1/2}$.

The nucleon source is modulated at a frequency of 0.84\,Hz to generate the exotic field, however, spurious signals can also be generated by the modulation of the heavy-mass material.  
Therefore, efforts are made to isolate possible crosstalk between the nucleon source and the comagnetometer. (1) Vibrational coupling of the nucleon source is reduced by optimizing the parameters in the servo motor, and further isolated by placing the test mass and the comagnetometer on separate foundations and additionally to mount the comagnetometer on a vibration-isolation platform. An aluminum  enclosure around the nucleon source is used to shield acoustic coupling through the air.
(2) Magnetic leakage from the tungsten and the  motor is measured with a fluxgate magnetometer. At the position near the magnetic shielding of the comagnetometer, the magnetic leakage is smaller than 0.1\,nT around 0.836\,Hz. 
After the decay with the distance and being shielded by five layers of mu-metal shield and one layer of ferrite shield, this field is estimated to be less than 1\,aT and therefore is negligible. (3) To diminish the motor's electromagnetic interference with the comagnetometer system, the comagnetometer is installed in a magnetic shielding room ($10\times 4 \times 3$\,m ), while the servo motor is controlled with a separate computer and the control units are placed outside the shielding room. The connection cable is shielded from radio-frequency noise.

\subsection{Data Analysis}
The averages of $S_i^{exp}$ are taken over the same intervals as $\bar{S}_{2i,+}^{sim}$ and $\bar{S}_{2i,-}^{sim}$, 
and they are read as $\bar{S}^{exp}_{2i,+}$,and $\bar{S}^{exp}_{2i,-}$, respectively.
Assuming that the background signal is time dependent, 
i.e. $n(t)=a+b*t+c*t^2$, where a, b and c are constants,
the exotic-force signals can be obtained by
\begin{equation}
\bar{S}_{i}=\frac{1}{8}\left[\bar{S}_{2i,+}-3\bar{S}_{2i,-}+3\bar{S}_{2i+1,+}-\bar{S}_{2i+1,-}\right],
\label{eq.Quad_BG}
\end{equation} 
where $\bar{S}_{i}$ represents $\bar{S}_{i}^{sim}$ or $\bar{S}_{i}^{exp}$. This operation removes frequency drifts that scale as $t^n$ with $n=0,1,2$.

\begin{acknowledgments}
We thank Xing Heng and Zitong Xu for debugging and operating the servo motor.  
 This work is supported by the National Natural Science Foundation of China (NSFC) (Grant Nos. 61925301 for Distinguished Young Scholars and 11875191),
the China postdoctoral Science Foundation (Grant No. 2021M700345), the DFG Project ID 390831469: EXC 2118 (PRISMA+ Cluster of Excellence), by the German Federal
Ministry of Education and Research (BMBF) within the Quantumtechnologien program (Grant No. 13N15064).
and by the QuantERA project LEMAQUME (DFG Project Number 500314265).

Author contributions: 
K.W., W.J., C.B.F, J.C.F., and D.B. proposed this study.  K.W. and W.J. performed the experiment and analyzed the data. 
K.W., W.J., C.B.F., A. W., V. F. and D.B. wrote the manuscript. 

Competing interests: The authors declare that they have no competing interests. 

Data and materials availability: All data needed to evaluate the conclusions in the paper are present in the paper. Additional data related to this paper can be requested from the authors.

\end{acknowledgments}

\bibliography{5thForce}

\end{document}